\documentclass[a4paper,11pt]{article}
\usepackage{jheppub}
\preprint{}% for details on the use of the package, please see the JINST-author-manual
\usepackage{lineno}
\usepackage{amsmath,amssymb}
\usepackage{physics}
\usepackage{parskip}
\usepackage{enumitem}
\usepackage{xcolor}

\definecolor{pink1}{rgb}{0.858, 0.188, 0.478}

\def\beq{\begin{equation}}
\def\eeq{\end{equation}}

\title{\boldmath Time Dependent String Compactification: Towards Bouncing Cosmology}

\author{Mir Mehedi Faruk$^{a,b}$,}\affiliation[a]{
Center for Fundamental Physics, School of Physical Science and Technology,
ShanghaiTech University, 393 Middle Huaxia Road, Shanghai 201210, China.
} 
\affiliation[b]{Department of Physics, McGill University, Montreal, QC, H3A 2T8, Canada.} 
\emailAdd{mmfaruk@shanghaitech.edu.cn,mir.faruk@mail.mcgill.ca, muturza3.1416@gmail.com,}

\abstract{We study the Null Energy Condition (NEC) arising from the Virasoro constraint on the string worldsheet. We then analyze how the NEC in the external spacetime directions emerges under general time-dependent string compactifications. Finally, we exhibit compactifications in which the averaged Einstein-frame condition allows the lower-dimensional description of the external spacetime to violate the NEC, thereby realizing a bouncing cosmology, while the higher-dimensional NEC remains satisfied, as dictated by worldsheet symmetry. We comment on scale-separated solutions obtained through the averaged Einstein-frame condition.}

\begin{document}
\maketitle

\flushbottom
\Large
\section{Introduction}\large Our universe exhibits cosmological acceleration in both its early and late-time history, motivating the search for accelerating solutions within quantum-gravity frameworks such as string theory\cite{Kachru:2003aw,Witten:2001kn,Banks:2000fe,Goheer:2002vf,Montefalcone:2020vlu,Cornalba:2002nv}. A natural question is whether the 10/11-dimensional supergravity theories that arise as low-energy effective field theories of string/M-theory permit such cosmologies\cite{Dasgupta:2019vjn,Kounnas:2011gz,Russo:2019fnk,Faruk:2024usu,Bernardo:2021zxo,Obied:2018sgi,Steinhardt:2008nk,Townsend:2003fx,Shiu:2023nph,Hartong:2006rt}. Realizing accelerating geometries in this setting is obstructed by a number of no-go theorems, most notably the Gibbons--Maldacena--Nu\~nez (GMN) theorem\cite{Maldacena:2000mw,Gibbons:2003gb} and its descendants\cite{VanRiet:2023pnx,Bernardo:2022ony,Niedermann:2025tuo,Dasgupta:2018rtp,Green:2011cn,Kutasov:2015eba,Kamal:2025qia,Shiu:2023fhb,Ohta:2004wk}, which forbid compactification of the $D$-dimensional Einstein equations to a $d$-dimensional $(D>d)$, positively curved, accelerating cosmology via a time-independent compactification, provided the $D$-dimensional stress tensor satisfies the strong energy condition\cite{Dasgupta:2014pma,Lehnert:2025izp,Koster:2011xg}. Within the two-derivative approximation, classical string-theoretic ingredients are insufficient to evade GMN, fueling ongoing debate on the status of accelerating cosmologies in string theory under the swampland program\cite{Palti:2019pca,Lehnert:2025izp}.\\\\
Energy conditions are central to the structure and physical implications of general relativity\cite{Curiel:2014zba,Barcelo:2002bv,Tipler:1978zz,Kontou:2020bta,Parikh:2014mja}, and many foundational results rely on assuming that certain energy conditions hold. We focus in this paper on the Null Energy Condition (NEC), one of the most robust energy conditions and a crucial ingredient of the Penrose singularity theorems\cite{Penrose:1964wq}, which is satisfied by many effective theories emerging from string theory and appears in a range of no-go arguments ruling out, for example, cosmological bounces\cite{Brandenberger:1993ef,Bernardo:2021zxo,Parikh:2015bja}. For closed bosonic strings propagating in a curved geometry, the Virasoro constraint precisely gives rise to the NEC to leading order in the $\alpha'$ expansion~\cite{Parikh:2015wae,Parikh:2014mja,Parikh:2015bja,Chatterjee:2015uya}, so the worldsheet symmetry enforces the NEC in higher dimensional spacetime $M_D$. For a compactification with direct product topology $M_D=\bar{M}_{d} \times \mathcal{M}_{D-d}$, where $\mathcal{M}$ denotes the compact directions and $\bar{M}$ the external directions\cite{Parikh:2014mja}, in the time-independent case the NEC in the external directions $\bar{M}_{d}$ is inherited from the higher-dimensional geometry $M_D$\cite{Parikh:2014mja,Bernardo:2022ony,Parikh:2015bja,Bernardo:2021zxo}, with significant consequences for early universe cosmology~\cite{Carroll:2004st,Battefeld:2014uga}. An immediate corollary is that open or flat bouncing cosmological solutions, which require NEC violation in four-dimensional spacetime~\cite{Chatterjee:2015uya,Parikh:2015bja}, are not consistent with string theory in this setting.\\\\
Time-dependent compactification provides a clear way to circumvent many of the no-go theorems arising in string cosmology, including those tied to EFT validity, scale separation, flux equations of motion, and entropy bounds\cite{Andriot:2025cyi,Gautason:2015tig,Emparan:2003gg,Coudarchet:2023mfs,Russo:2019fnk,VanRiet:2023pnx,Bernardo:2022ony,Niedermann:2025tuo,Dasgupta:2018rtp,Kounnas:2011gz,Faruk:2023uzs,Ohta:2003pu,Ishino:2005ru}. This is particularly relevant for bouncing cosmologies, which are often prescribed as a way to circumvent the Big Bang singularity --- the FLRW scale factor acquires a nonzero minimum, with a contracting phase preceding the current expansionary era\cite{Novello:2008ra,Ijjas:2018qbo} --- but require NEC violation and are therefore problematic to construct in string theory\cite{Parikh:2015wae}. Such scenarios can be accommodated using time-dependent compactification\cite{Russo:2019fnk,Novello:2008ra,Battefeld:2014uga,Parikh:2015bja}, as we discuss in detail in this article.\\\\
The paper is organized as follows. 
The worldsheet symmetry through Virasoro constraints that lead to the NEC in higher dimensions includes the metric and the dilaton\cite{Parikh:2014mja,Parikh:2015wae,Parikh:2015bja,Chatterjee:2015uya}, but omits the Kalb--Ramond $B$ field. The fundamental strings in string theory are charged under such antisymmetric field, therefore of fundamental importance\cite{Becker:2006dvp,Green:1987sp}.
We first extend the worldsheet arguments more robustly by including the $B$ field,  in Section~\ref{v}. Then, in Section~\ref{section:unaveraged}, we turn our attention to time-dependent compactification and identify the constraints that yield solutions which, upon compactification, satisfy the NEC in the external directions. Finally, we discuss the averaged condition that arises from imposing the Einstein-frame condition when integrating over the compact space in Section~\ref{bq}.
We systematically investigate how the NEC in external spacetime directions emerges in general time-dependent string compactifications.
In this class of compactifications, it is possible to find solutions that violate the NEC in the external directions but satisfy the NEC in higher dimensions, as dictated by worldsheet symmetry\cite{Parikh:2014mja}. As a result, bouncing cosmological solutions can fit within such a compactification scheme. We present one such example in Section~\ref{5}. We make an interesting observation on scale separated solutions. We finish the article with discussion and future works on the application of time-dependent compactification on string theory configurations.
\section{NEC from string worldsheet}
\label{v}
The most
general action describing a string coupled with background metric $g_{MN}$, dilaton $\Phi$ and Kalb--Ramond 
field $B_{MN}$
is given by,\footnote{$a,b$ and $M,N$ are worldsheet and spacetime indices}
\begin{eqnarray}
S=\frac{1}{4\pi \alpha'}\int d^2\sigma
\sqrt{h} \Bigg[\bigg(h^{ab} g_{MN}(X)+i\epsilon^{ab}B_{MN}(X)\bigg)
\partial_a X^M\partial_bX^N
+\alpha' R \Phi(X)\Bigg]
\end{eqnarray}
The stress tensor can receive contributions from three different fields,
\begin{eqnarray}
    <T^P_P>=-\frac{1}{2\alpha'}\beta_{MN}(G) g^{ab}\partial_a
   X^M \partial_b X^N
   -\frac{i}{2\alpha'}\beta_{MN}(B)\epsilon^{ab}\partial_aX^M\partial_bX^N
-\frac{1}{2}   \beta(\Phi) R\nonumber\\
\end{eqnarray}
To first order in $\alpha'$, the beta functions are given by (in string frame)\cite{Becker:2006dvp}:
\begin{eqnarray}
   && \beta_{MN}^G=\alpha'R_{MN}+2\alpha'\nabla_M\nabla_N\Phi-
    \frac{\alpha'}{4}H_{MPQ}H_N^{PQ}\\
     &&\beta_{MN}^B=-\frac{\alpha'}{2}\nabla^P 
    H_{P
     MN}
     +{\alpha'}\nabla^P\Phi H_{P MN}\\
&&     \beta^\Phi=-\frac{\alpha'}{2}\nabla^2\Phi+\alpha'\nabla_M \Phi\nabla^M\Phi-\frac{\alpha'}{24}H_{PQL}H^{PQL}
\end{eqnarray}
One key feature of a conformal field theory is its traceless property of the stress-tensor due to Weyl invariance.
A consistent background of string theory must preserve this traceless property\cite{Becker:2006dvp} also known as Virasoro constraint conditions. This can be implied by, $\beta^G_{MN}=\beta^B_{MN}=\beta^\Phi=0$. The vanishing beta function for the metric implies that,
\begin{eqnarray}
  (R_{MN}+ 2\nabla_M\nabla_N\Phi-\frac{1}{4}H_{MPQ}H_{N}^{P
    Q})
    =0
\end{eqnarray}
We are going to study the above equation in Einstein frame, where the metric is related to the string-frame metric as below,
\begin{eqnarray}
       g_{MN}^E=e^{-\frac{4\Phi}{D-2}}g_{MN}  
\end{eqnarray}
The Ricci tensor in Einstein frame is related to the Ricci tensor in string frame through the following relation,
\begin{align} 
R^{E}_{MN}
=
R_{MN}
+2\nabla_M\nabla_N\Phi
+\frac{2}{D-2}\,g_{MN}\nabla^{2}\Phi
+\frac{4}{D-2}\,\nabla_M\Phi\,\nabla_N\Phi
-\frac{4}{D-2}\,g_{MN}(\nabla\Phi)^2
\,.
\end{align}
Once again, for the vanishing dilaton beta function (first order in $\alpha'$),
\begin{eqnarray}
   \nabla^2\Phi=2\nabla^M\Phi\nabla_M\Phi-H^2/12\,,
\end{eqnarray} We can now write the Ricci tensor in the Einstein frame only in terms of first derivatives of the dilaton and the $B$-field.
\begin{align} 
R_{MN}^E&=\frac{1}{4}H_{MPQ}H_N^{~PQ}+\left(\frac{4}{D-2}\right)(\nabla_M\Phi)(\nabla_N\Phi)\nonumber\\ 
&\quad -g_{MN}\left(\left(-\frac{2}{D-2}\right)\left(2\nabla^Q\Phi\nabla_Q\Phi-\frac{H^2}{12}\right)+(D-2)\left(-\frac{2}{D-2}\right)^2(\nabla_Q\Phi)(\nabla^Q \Phi)\right)\nonumber\\
&=\left(\frac{4}{D-2}\right)(\nabla_M\Phi)(\nabla_N\Phi)+\frac{1}{4}H_{MPQ}H_N^{~PQ}-\frac{1}{6(D-2)}g_{MN}H^2\nonumber\\
&=\left(\frac{4}{D-2}\right)(\nabla_M^E\Phi)(\nabla_N^E\Phi)+\left(\frac{1}{4}H_{MPQ}\tilde{H}_N^{~PQ}-\frac{1}{6(D-2)}g_{MN}^E\tilde{H}^2 \right)e^{-8\Phi/(D-2)}\,,\label{xd}
\end{align}
We can now further examine the {NEC} from \eqref{xd}. Note that,
 in the last line we wrote the RHS using the metric in the Einstein frame to raise the indices of $H_{MNP}$ (we denoted this by $\tilde{H}$). Contracting the Ricci tensor in the Einstein frame with null vectors $l^M$ (i.e. $l^2=l^M l^N g_{MN}=0$) yields 
\begin{eqnarray}
R_{MN}^E l^M l^N &=&\left(\frac{4}{D-2}\right)(l^M\nabla_M\Phi)^2+\frac{1}{4}\overbrace{l^M l^N H_{MPQ}H_N^{~PQ} }^{\equiv C_B}\nonumber\\
&=&\left(\frac{4}{D-2}\right)(l^M\nabla_M^E\Phi)^2+\left(\frac{1}{4}l^M l^N H_{MPQ}\tilde{H}_N^{~PQ} \right)e^{-8\Phi/(D-2)}\,,
\end{eqnarray}
where in the first and second lines, the RHS is written in the string and Einstein frames, respectively. Notice that the first term of the RHS is manifestly non-negative, and we will show in the next section that the second term $C_B/4\geq 0\iff C_B\geq 0$ is also non-negative for any null vector $l^M$, which implies the NEC:
\beq 
R_{MN}^E l^M l^N\geq 0\,~~\forall~~l^M~~\text{such that }l^2=0\,.\label{abdul}
\eeq 
We would like to emphasize that this result is expected, since previous work on worldsheet symmetry~\cite{Parikh:2014mja,Parikh:2015wae} as well as the low-energy effective description of string theory are well known to satisfy the higher-dimensional energy condition~\cite{Bernardo:2022ony}. However, we stress that the NEC-satisfying property of the higher-dimensional theory makes it more difficult to violate the NEC in the external directions. This can be achieved using time-dependent string compactification.

\subsection{$
C_B\geq 0
$}
\label{2.1}
In this section, we show that the contraction
\beq 
C_B\equiv l^M l^N H_{MPQ}H_{NP'Q'}g^{PP'}g^{QQ'}\,,
\eeq 
involving the field strength $H=dB$ of the Kalb--Ramond field $B$, and an arbitrary null vector $l^M$, is always non-negative. It is convenient to go to a local vielbein basis $e_{M}^{~a}(x)$ (see Appendix G of \cite{Carroll:2004st} for its definition and properties) by writing any tensor as
\beq 
T_{M}^{~N}= e_{M}^{~a} e^{N}_{~b} T_a^{~b}\iff T_a^{~b}= e^{M}_{~a} e_{N}^{~b} T_{M}^{~N}\,,
\eeq 
where we specialized to a $(1,1)$ tensor for simplicity, and the metric as
\beq 
g^{PQ}=e^{P}_{~a} e^{Q}_{~b}\eta^{ab}\,, 
\eeq
where $\eta^{ab}$ is the Minkowski metric. This allows us to write the contraction $C_B$ as
\beq\label{CB-abindices}
C_B\equiv l^a l^b H_{acd}H_{bef}\eta^{ce}\eta^{df}\,.
\eeq 
Instead of taking the vielbein $e_a^{~M}$ to be such that $\eta_{ab}=g_{MN}e_a^{~M}e_b^{~N}=\text{diag}(-1,1,\cdots,1)$ is the usual Minkowski metric that corresponds to the cartesian coordinates in flat space, we will change variables to a new vielbein
\beq 
e_{~\tilde{a}}^{M}=C_{\tilde{a}}^{~a}e_{~a}^{M}\,,
\eeq  
such that the flat metric is
\beq \label{eq-metric-abtilde}
\eta_{\tilde{a}\tilde{b}}=g_{MN}e_{~\tilde{a}}^{M}e_{~\tilde{b}}^{N}=C_{\tilde{a}}^{~a}C_{\tilde{b}}^{~b}\eta_{ab}=\begin{pmatrix}\begin{matrix}0&-1/2\\-1/2&0\end{matrix}&0 \\
0&\delta_{ij}\end{pmatrix}\,,
\eeq 
which corresponds to the null coordinates $(u,v,x^i)$ of flat space, where the metric is $ds^2=-du\, dv+dx^i dx^i$ (with $i=2,3,\cdots,D-1$). This new vielbein $e_{~\tilde{a}}^{M}$ is suited to work with null vectors $l^M=e_{~\tilde{a}}^{M}l^{\tilde{a}}$, because we can always perform a local Lorentz transformation to write the null vector $l^{\tilde{a}}$ as
\beq\label{eq-nullvector-tildea}
l^{\tilde{a}}=l^u(1,0,\cdots,0)\,,
\eeq 
where we denote the components of the index $\tilde{a}$ as $V^{\tilde{a}}=(V^u,V^v,V^i)$. Indeed, this vector $l^{\tilde{a}}$ is manifestly null with the metric \eqref{eq-metric-abtilde}.\\\\
Coming back to the contraction $C_B$ in equation \eqref{CB-abindices}, we can write it using the new vielbein basis $e_{~\tilde{a}}^{M}$ as
\begin{align}\label{CB-abindices1}
C_B&\equiv l^{\tilde{a}} l^{\tilde{b}} H_{\tilde{a}\tilde{c}\tilde{d}}H_{\tilde{b}\tilde{e}\tilde{f}}\eta^{\tilde{c}\tilde{e}}\eta^{\tilde{d}\tilde{f}}\nonumber\\
&=(l^u)^2 H_{u\tilde{c}\tilde{d}}H_{u\tilde{e}\tilde{f}}\eta^{\tilde{c}\tilde{e}}\eta^{\tilde{d}\tilde{f}}\,,
\end{align}
where in the second line we chose the frame such that the null vector $l^{\tilde{a}}$ has the form of equation \eqref{eq-nullvector-tildea}. Notice that the only non-trivial contribution in the index sum of the second line is when the indices $\tilde{c}$, $\tilde{d}$, $\tilde{e}$, $\tilde{f}$ that we are summing over take the values $i=2,3,\cdots,D-1$ \footnote{Firstly, there is trivially no contribution from taking one of these indices, for example $\tilde{c}$, to be $\tilde{c}=u$ because $H_{uu\tilde{d}}=0$ as $H$ is an antisymmetric tensor. Secondly, there is also no contribution from taking one of these indices, for example $\tilde{c}$, to be $\tilde{c}=v$ because $H_{uv\tilde{d}}H_{u\tilde{e}\tilde{f}}\eta^{v\tilde{e}}=H_{uv\tilde{d}}H_{uu\tilde{f}}\eta^{vu}=0$ where we used that $\eta^{v\tilde{e}}$ is non-zero only if $\tilde{e}=u$.}. Therefore, since the metric $\eta^{ij}=\delta^{ij}$ is simply the Kronecker delta for the indices $i=2,3,\cdots,D-1$, we can write the contraction as
\begin{align}\label{CB-abindices2}
C_B&=(l^u)^2 H_{uik}H_{ujl}\delta^{ij}\delta^{kl}\geq 0\,,
\end{align}
which is manifestly non-negative, as we wanted to show. We will now proceed to show the inheritance of the NEC in lower dimensions.
\section{String compactification and NEC I}
\label{section:unaveraged}
We start with the following general warped (time-dependent) compactification. 
The metric is chosen to be in Einstein frame\cite{Bernardo:2022ony},
\begin{equation}\label{eq-metric}
    d{s}^2 ={g}^E_{MN}(x,y)dx^M dx^N = \Omega^2(y)\bar{g}_{\mu\nu}(x) dx^\mu dx^\nu + h_{mn}(x,y)dy^m dy^n,
\end{equation}
Here, the $\bar{g}_{\mu\nu}$ is the metric on external spacetime of dimension $d$ and $h_{mn}$ is a symmetric positive definite metric over compact internal space of dimension $n=D-d$ and $\Omega(y)$ is the smooth warp function.
The Ricci tensor of the external direction is related to the unwarped Ricci tensor by~\cite{Bernardo:2022ony},
\begin{eqnarray}
    {R}_{\alpha\beta}^E &= \bar{R}_{\alpha\beta}(\bar{g})-\frac{1}{4}\bar{\nabla}_{\alpha}h^{pq}\bar{\nabla}_\beta h_{pq} -\frac{1}{2}h^{pq}\bar{\nabla}_\alpha\bar{\nabla}_\beta h_{pq} -\frac{1}{d}\bar{g}_{\alpha\beta}\Omega^{2-d}\nabla^2\Omega^d
\end{eqnarray}\large
In Einstein
frame, the effective $d$-dimensional gravity theory has the following Newton's constant:
\begin{eqnarray}
    \frac{1}{G_d}=\frac{1}{G_D}\int d^n {y}\,  \Omega^{d-2} \, \sqrt{\det h}\label{pab}
\end{eqnarray}
$G_d$ and $G_D$ are Newton's constant in $d$ and $D$ dimensions respectively.
The $d$-dimensional Newton's constant $G_d$ is observed to be time independent. 
As a result, the time independence of the internal volume is related to the time independence of Newton's constant in $d$ dimensions.
Therefore, 
imposing the internal volume to be external space-time $x^\mu$ independent,
\begin{equation}
    \partial_\mu \sqrt{\, \det h} = 0 \label{nh} ,
\end{equation} 
so that four dimensional Newton's constant $G_d$ does not vary with space and time~\eqref{pab}.
Imposing the internal volume to be time independent (it is also going to be  
independent
of external
space\footnote{denoted by the $\mu$ index}) we find, 
\begin{equation}
\label{constvol}
    \nabla_\mu \sqrt{\text{det}h} = 0 \implies h^{mn}\nabla_\mu h_{mn}=0\,, 
    \end{equation}
    %This enables us to conclude,
%\begin{eqnarray}
%    \quad \nabla_M h^{mn}\nabla_\mu h_{mn} = - h^{mn}\nabla_M \nabla_\mu h_{mn}\,.\nonumber
%\end{eqnarray}
    Therefore,
with such a constraint the Ricci tensor of the external direction is related to the unwarped Ricci tensor by
\begin{eqnarray}
    {R}^E_{\mu\nu} = \bar{R}_{\mu\nu}(\bar{g}) +\frac{1}{4} \bar{\nabla}_\mu h^{pq}\bar{\nabla}_\nu h_{pq} - \frac{1}{d}\bar{g}_{\mu\nu} \Omega^{2-d}\nabla^2 \Omega^d\,,\label{kakku}
\end{eqnarray}
The $D$-dimensional  NEC in Einstein frame is given by
\begin{equation}
    {R}_{MN}^El^Ml^N \geq 0, \quad {g}_{MN}l^{M}l^{N}=0. \nonumber
\end{equation}
When taking the null 
vectors in the external directions, eq.~\eqref{kakku}
dictates that
\begin{eqnarray}
    {R}^E_{\mu\nu}l^\mu l^\nu = \bar{R}_{\mu\nu}(\bar{g})l^\mu l^\nu +\frac{1}{4} \bar{\nabla}_\mu h^{pq}\bar{\nabla}_\nu h_{pq} l^\mu l^\nu ,\label{kakku1}
\end{eqnarray}
The last term in eq.~\eqref{kakku} vanishes because of null vectors.
But we have the following identity:
\begin{equation}
    \bar{\nabla}_\mu h^{pq} = - h^{pm}h^{qn}
    \bar{\nabla}_\mu h_{mn},
\end{equation}
As a result, 
\begin{equation}
    \bar{R}_{\alpha\beta}(\bar{g})l^\alpha l^\beta \geq +\frac{1}{4} h^{mp}(l^\alpha\bar{\nabla}_\alpha h_{mn})h^{nq}(l^\beta\bar{\nabla}_\beta h_{pq})= \frac{1}{4}\text{tr}\left[h^{-1}(l\cdot \bar{\nabla}h)\right]^2 \geq 0,\label{bacchu}
\end{equation}
which is the $d$-dimensional NEC for the metric $\bar{g}$, i.e. the external metric whose NEC is of interest to us.
The internal metric is not assumed to be diagonal and therefore the last inequality of \eqref{bacchu} is not straightforward.
This inequality in the last step of \eqref{bacchu} is proved in Appendix \ref{app-proof-trace}. We should of course note that when \eqref{nh} is not satisfied, then the 4D NEC is not guaranteed even though the higher-dimensional theory respects the NEC. However, not maintaining \eqref{nh} would also indicate that Newton’s constant is time-dependent.
%\\\\\\\\
\section{String compactification and NEC II: A more general case}
\label{bq}
Once again starting from \eqref{eq-metric},
with general time-dependent compact internal manifold,
\begin{equation}\label{eq-metric-Omega-t}
    d{s}^2 ={g}^E_{MN}(x,y)dx^M dx^N = \underbrace{\Omega^2(x,y)\bar{g}_{\mu\nu}(x)}_{\displaystyle\equiv g_{\mu\nu}(x,y)} dx^\mu dx^\nu + h_{mn}(x,y)dy^m dy^n,
\end{equation}
Here, $\bar{g}_{\mu\nu}$ is the metric on external spacetime of dimension $d$ without the warp factor $\Omega^2(x,y)$, while $g_{\mu\nu}\equiv \Omega^2\bar{g}_{\mu\nu}$ is the metric on the external spacetime with the warp factor, and $h_{mn}$ is a symmetric positive definite metric over compact internal space of dimension $n=D-d$.

\subsection{Constraint from Newton's constant}
Recalling that the effective $d$-dimensional gravity theory has the following Newton's constant,
\begin{eqnarray}
    \frac{1}{G_d}=\frac{1}{G_D}\int d^n {y}\,  \Omega^{d-2} \, \sqrt{\det h}\label{pab-2}.
\end{eqnarray}
Sometimes referred to as the Einstein-frame condition.
To consider solutions with constant Newton's constant $G_d$, we take the derivative with respect to the external directions of the above equation \eqref{pab-2}\cite{Russo:2019fnk}.

\beq \label{eq-intXmu}
\int d^n {y}\,  \Omega^{d-2} \, \sqrt{\det h} ~ X_\mu =0\,,
\eeq 
where we defined
\beq \label{eq-def-Xmu}
X_\mu\equiv \bar{\nabla}_\mu \left(\ln\left(  \Omega^{d-2} \, \sqrt{\det h}\right)\right)=\frac{1}{2}\text{tr}\left(h^{-1}\bar{\nabla}_\mu h\right)+(d-2)\frac{\bar{\nabla}_\mu\Omega }{\Omega}\,.
\eeq 
Therefore, there are two different ways of achieving constant $G_d$,\\
1) A trivial way by setting the \textbf{unaveraged condition}, i.e.  $X_\mu=0.$ This will give us back the result of section \ref{section:unaveraged} through \eqref{constvol}. That means it will make the volume of internal space time (and space) independent. Therefore such compactification will give rise to solution which satisfies NEC.\\\\
2) On the other hand, we can go for a more involved condition, the \textbf{averaged condition}, i.e. $<X_\mu>=0$. This means, we impose \eqref{eq-intXmu} only after integrating over the compact manifold.
\\\\
Taking the covariant derivative of \eqref{eq-intXmu} further implies\cite{Steinhardt:2008nk,Russo:2018akp}:
\beq \label{eq-intnablaXmu}
\int d^n {y}\,  \Omega^{d-2} \, \sqrt{\det h} ~\left(\bar{\nabla}_\nu X_\mu+X_\mu X_\nu\right) =0\,,
\eeq 
where 
\beq 
\bar{\nabla}_\nu X_\mu=\frac{1}{2}\text{tr}(h^{-1}\bar{\nabla}_\nu\bar{\nabla}_\mu h)+(d-2)\frac{\bar{\nabla}_\nu\bar{\nabla}_\mu\Omega }{\Omega}+\frac{1}{2}\bar{\nabla}_\nu h^{pq} \bar{\nabla}_\mu h_{pq}-(d-2)\frac{\bar{\nabla}_\nu\Omega }{\Omega}\frac{\bar{\nabla}_\mu\Omega }{\Omega}\,.
\eeq 
\subsection{Towards understanding the four dimensional NEC}
The Ricci tensor of the external direction is related to the unwarped Ricci tensor by
\begin{align}\label{eq-Ricci-Omegagbar}
    {R}_{\alpha\beta}(g^E) &= R_{\alpha\beta}(g)-\frac{1}{4}\nabla_{\alpha}h^{pq}\nabla_\beta h_{pq} -\frac{1}{2}h^{pq}\nabla_\alpha\nabla_\beta h_{pq}
\nonumber\\
&\quad\quad +\frac{1}{2}g^{\rho\sigma}\nabla_pg_{\rho\alpha}\nabla^pg_{\sigma\beta}-\frac{1}{4}g^{\rho\sigma}\nabla_pg_{\alpha\beta}\nabla^pg_{\rho\sigma}-\frac{1}{2}h^{pq}\nabla_p\nabla_qg_{\alpha\beta}\,,
\end{align}
where the covariant derivatives $\nabla_\alpha$ are with respect to the $d$-dimensional metric $g_{\mu\nu}(x,y)\equiv\Omega^2(x,y)\bar{g}_{\mu\nu}(x)$, and the covariant derivatives are with respect to the $n$-dimensional metric $h_{mn}$.
It is possible to write the above expression in terms of the metric $\bar{g}_{\mu\nu}(x)$ using,
\begin{align}
R_{\alpha\beta}(g)&=R_{\alpha\beta}(\bar{g})-\left((d-2)\frac{\bar{\nabla}_\alpha \bar{\nabla}_\beta \Omega}{\Omega}+\bar{g}_{\alpha\beta} \frac{\bar{\nabla}^\sigma\bar{\nabla}_\sigma\Omega}{\Omega}\right)\nonumber\\
&\quad \quad+2(d-2)\frac{\bar{\nabla}_\alpha \Omega}{\Omega}\frac{\bar{\nabla}_\beta \Omega}{\Omega}-(d-3)\bar{g}_{\alpha\beta} \frac{\bar{\nabla}^\sigma \Omega}{\Omega}\frac{\bar{\nabla}_\sigma \Omega}{\Omega}\,.
\end{align}
Finally, the \eqref{eq-Ricci-Omegagbar}
can be written in the following form ,
\begin{align}\label{eq-Ricci-gbar-withX}
{R}_{\alpha\beta}(g^E) &= R_{\alpha\beta}(\bar{g})-(d-2)\frac{\bar{\nabla}_\alpha \Omega}{\Omega}\frac{\bar{\nabla}_\beta \Omega}{\Omega}-\frac{1}{4}\text{tr}\left(h^{-1}(\bar{\nabla}_\alpha h )h^{-1}\bar{\nabla}_\beta h\right)
+\frac{\bar{\nabla}_\alpha\Omega}{\Omega}X_\beta+\frac{\bar{\nabla}_\beta\Omega}{\Omega}X_\alpha
\nonumber\\
&-\bar{\nabla}_\beta X_\alpha-\bar{g}_{\alpha\beta} \bigg(\frac{\bar{\nabla}^\sigma\bar{\nabla}_\sigma\Omega}{\Omega}+(d-3) \frac{\bar{\nabla}^\sigma \Omega}{\Omega}\frac{\bar{\nabla}_\sigma \Omega}{\Omega}+\frac{1}{2}\frac{\bar{\nabla}^\sigma\Omega}{\Omega}\text{tr}(h^{-1}\bar{\nabla}_\sigma h)
+\frac{1}{d}\Omega^{2-d}\nabla^p\nabla_p\Omega^d\bigg)\,.
\end{align}
Contracting this equation with null vectors $l^\alpha l^\beta$ where $l^2=\bar{g}_{\alpha\beta} l^\alpha l^\beta=0$, the terms of the third and fourth lines vanish, and we obtain,\footnote{See Appendix \ref{appendix B} for detailed computation}
\begin{align}\label{eq-Ricci-ll-withX}
{R}_{\alpha\beta}(g^E)l^\alpha l^\beta &=l^\alpha l^\beta R_{\alpha\beta}(\bar{g})-(d-2)\frac{\bar{\nabla}_l \Omega}{\Omega}\frac{\bar{\nabla}_l \Omega}{\Omega}-\frac{1}{4}\text{tr}\left(h^{-1}(\bar{\nabla}_l h )h^{-1}\bar{\nabla}_l h\right)\nonumber\\
&\quad\quad+2\frac{\bar{\nabla}_l\Omega}{\Omega}X_l-\bar{\nabla}_l X_l\,,
\end{align} where we defined $\bar{\nabla}_l\equiv l^\alpha \bar{\nabla}_\alpha$ and $X_l\equiv l^\alpha X_\alpha$.
The above identity in eq. \eqref{eq-Ricci-ll-withX}
is quite useful in understanding 
the unaveraged and averaged constraints on Newton's constant.
\footnote{One can easily note that unaveraged condition $X_\mu=0$ easily gives back the result of section \ref{section:unaveraged}.  Rewrite eq. \eqref{eq-Ricci-ll-withX} for unaveraged case,
\begin{align}\label{eq-Ricci-ll}
l^\alpha l^\beta R_{\alpha\beta}(\bar{g})&=l^\alpha l^\beta{R}_{\alpha\beta}(g^E) +(d-2)\left(\frac{\bar{\nabla}_l \Omega}{\Omega}\right)^2+\frac{1}{4}\text{tr}\left[(h^{-1}(\bar{\nabla}_lh))^2\right]\nonumber\\
&\geq l^\alpha l^\beta{R}_{\alpha\beta}(g^E)\,,
\end{align}
where in the second line we used that $\left(\bar{\nabla}_l \Omega/\Omega\right)^2\geq 0$ and that $\text{tr}\left[(h^{-1}(\bar{\nabla}_lh))^2\right]\geq 0$ (as proved in Appendix \ref{app-proof-trace}). If we assume the higher-dimensional NEC condition $l^\alpha l^\beta{R}_{\alpha\beta}(g^E) \geq 0$, then we can see that this trivially implies the lower-dimensional NEC
\beq 
l^\alpha l^\beta R_{\alpha\beta}(\bar{g})\geq0\,,
\eeq 
This is the reason behind the NEC inheritance in the previous section \eqref{bacchu}, meaning the lower dimensional NEC is inherited from higher dimensions. We will now see in 
section
\ref{chapasec}, that it is not the case for the averaged condition
in general.} 
\subsubsection{Averaged condition}
\label{chapasec}
Starting from  \eqref{eq-Ricci-ll-withX}, and
integrating over the compact internal manifold,
\begin{eqnarray}
    \int d^n {y}\,  \Omega^{d-2} \, \sqrt{\det h}
  {R}_{\alpha\beta}(g^E)  l^\alpha l^\beta && =\int d^n {y}\,  \Omega^{d-2} \, \sqrt{\det h}\big(-(d-2)\frac{\bar{\nabla}_l \Omega}{\Omega}\frac{\bar{\nabla}_l \Omega}{\Omega} -\frac{1}{4}\text{tr}\left(h^{-1}(\bar{\nabla}_l h )h^{-1}\bar{\nabla}_l h\right)\nonumber\\
&&+\frac{\bar{\nabla}_l\Omega}{\Omega}X_l+\frac{\bar{\nabla}_l\Omega}{\Omega}X_l-\bar{\nabla}_l X_l\big) 
+     \int d^n {y}\,  \Omega^{d-2} \, \sqrt{\det h}
  {R}_{\alpha\beta}(\bar{g})  l^\alpha l^\beta
\,, \nonumber\\\label{averagedfinal}
\end{eqnarray}
Now using eq.~\eqref{eq-intnablaXmu}, we can find that \eqref{averagedfinal} implies
\begin{eqnarray}\label{xxcfihmk}
    \int d^n {y}\,  \Omega^{d-2} \, \sqrt{\det h}
  {R}_{\alpha\beta}(\bar{g})  l^\alpha l^\beta 
=\int d^n {y}\,  \Omega^{d-2} \, \sqrt{\det h} \Big( {R}_{\alpha\beta}({g^E})  l^\alpha l^\beta +(d-1)(\frac{\bar{\nabla}_l \Omega}{\Omega})^2
\nonumber\\
+\frac{1}{4}\text{tr}\left[(h^{-1}(\bar{\nabla}_lh))^2\right]
-(X_l+\frac{\nabla_l \Omega}{\Omega})^2
\Big)\nonumber\\\label{AvgDhaka}
\end{eqnarray}
All terms in the RHS of \eqref{AvgDhaka} are positive except for the last term. 
The last term is a squared term, and therefore a negative semi-definite quantity. 
As a consequence, the left hand side is not guaranteed to be greater than zero. Therefore, the four dimensional spacetime is not guaranteed to satisfy the NEC although the higher dimensional spacetime satisfies the NEC.
We will now explore  interesting string compactification scenarios where  the external four dimensional NEC is violated giving rise to bouncing cosmology 
solutions despite the higher dimensional NEC is satisfied (as guided by worldsheet symmetry).
Before we move on, we note that considering the time-dependent internal manifold,
\begin{eqnarray}\label{eq:def-fk}
\Omega(t,y)\equiv e^{f(t,y)},\qquad \sqrt{\det h(t,y)}\equiv e^{k(t,y)}.
\end{eqnarray}
Then the Einstein-frame condition eq. (\ref{pab-2}) becomes
\begin{equation}
\frac{G_D}{G_d}=\int d^n y\, e^{A},
\end{equation}
with,
\begin{equation}
A(t,y)\equiv (d-2)f(t,y)+k(t,y).
\end{equation}
The Ricci tensor
becomes,
\begin{eqnarray}\label{eq:general-null-nullX-otp}
R_{\alpha\beta}(g_E)\,l^\alpha l^\beta
&= R_{\alpha\beta}(\bar g)\,l^\alpha l^\beta 
 + (l^{t})^{2}\Bigg[
-(d-2)\,\ddot f
+ (d-2)\,\dot f^{\,2}
+ 2\,\dot f\,\dot k
- \ddot k
-\frac{1}{4}\,\mathrm{tr}\!\left(\mathcal{B}^{2}\right)
\Bigg]\nonumber  \\
&\quad + l^{i}l^{j}\delta_{ij}\,a\dot a\,
\Big[(d-2)\,\dot f+\dot k\Big]\,,
\end{eqnarray}
where,
\begin{equation}\label{eq:def-B}
\mathcal{B}^{m}{}_{n}(t,y)\;\equiv\;(h^{-1}\dot h)^{m}{}_{n}
\;=\;h^{mp}(t,y)\,\dot h_{pn}(t,y),
\qquad
\mathrm{tr}(\mathcal{B}^{2})\;\equiv\;\mathcal{B}^{m}{}_{n}\,\mathcal{B}^{n}{}_{m}.
\end{equation}
\section{Bouncing Cosmology through Averaged Condition: An Example}
\label{5}
We now analyse the following metric in higher dimensions, in order to understand the interplay between averaged Einstein-frame condition, NEC and time-dependent compactification.
We take the external metric to be a flat 
FRW metric, the metric is given by
\begin{equation}
ds_D^2
=
\Omega^2(y;t)\, ds_{\mathrm{FRW}}^2
+
h_{mn}(t,y)\, dy^mdy^n \, .
\end{equation}
The internal manifold is taken to be of 
a compact manifold of size $L$ with
dimension $n$ of the following form,
\begin{eqnarray}
    h_{mn}(t,y)=e^{2u}(t,y)\hat{h}_{mn}(y)
\end{eqnarray}
Therefore, from \eqref{eq:general-null-nullX-otp} we have,
\begin{eqnarray}
     k \equiv \ln \sqrt{\det h} = \ln \sqrt{\det \hat h} + n u, \nonumber
%\item $\dot k = n \dot u$,%\item $\ddot k = n \ddot u$.
\end{eqnarray}

\noindent
The matrix:
\begin{equation}
\mathcal{B}^m{}_n
= (h^{-1}\dot h)^m{}_n
= 2\dot u\,\delta^m{}_n .
\end{equation}

\noindent
Hence:
\begin{equation}
\mathrm{tr}(\mathcal{B}^2)
= 4n\,\dot u^{\,2}
= \frac{4}{n}\,\dot k^{\,2} .
\end{equation}

\noindent
Therefore:
\begin{equation}
-\frac14\,\mathrm{tr}(\mathcal{B}^2)
= -\frac{1}{n}\,\dot k^{\,2} .
\end{equation}Thus, from \eqref{eq:general-null-nullX-otp},
\begin{equation}\label{soku}
R_E
=
R_{\bar g}
+
\Big[
-(d-2)\,\ddot f
+(d-2)\,\dot f^{\,2}
+2\,\dot f\,\dot k
-\ddot k
-\frac{1}{n}\,\dot k^{\,2}
\Big]
+
\frac{\dot a}{a}\,
\Big[(d-2)\,\dot f+\dot k\Big] .
\end{equation}
%-----------------------------------------------------------------
% Clean rewrite: Eqs. (6.6)--(6.9)
%-----------------------------------------------------------------

Let the internal space be an $n$-torus $T^n$ with equal radii $L$,
\[
y_i \sim y_i+2\pi L,\qquad i=1,\dots,n,
\]
and define
\[
\theta_i \equiv \frac{t+y_i}{L}\,.
\]
Then $y_i:0\to 2\pi L$ corresponds to $\theta_i:t/L\to t/L+2\pi$, and  all profiles below are $2\pi$-periodic. We can now integrate 
over internal compact manifold.
%each $\theta_i$ over $[0,2\pi]$.
We now take the following ansatz,
\begin{equation}\label{eq:ansatz-factorized-sin}
\Omega(t,\mathbf y)=2A\prod_{i=1}^n\bigl(1+pf(\theta_i)\bigr),
\qquad
\sqrt{\det h(t,\mathbf y)}=A\prod_{i=1}^n\bigl(1+2pg(\theta_i)\bigr),
\qquad (|p|<\tfrac12).
\end{equation}
where, $f$ and $g$ are periodic functions. Taking $f$ and $g$ to be $\sin(\theta)$\footnote{One can also choose for example, $f$ and $g$ to be $\cos{\theta}$ and the result in eq. \eqref{eq:REw-final-factorized-Tn}
will remain the same.
} and
setting $d=4$ and performing a straightforward simplification, the weighted average is,
\begin{equation}\label{eq:REw-final-factorized-Tn}
\boxed{
\langle R_E\rangle_w
=
R_{\bar g}(t)
+\frac{1}{2L^{2}(5p^{2}+2)}
\Big[
2(21n-3)p^{2}+(n-1)\bigl(1-\sqrt{1-4p^{2}}\bigr)
\Big],
\qquad (|p|<\tfrac12).
}
\end{equation}

We have defined the weighted average of any quantity $Z$ by,
\begin{equation}\label{eq:weighted_avg_clean}
\langle Z\rangle_{w}
\equiv
\frac{\displaystyle\int_{M_n} d^{n}y\; w(t,y)\,Z(t,y)}
{\displaystyle\int_{M_n} d^{n}y\; w(t,y)} \, .
\end{equation}where,\begin{equation}\label{eq:weight_clean}
w(t,y)\equiv \Omega^{d-2}(t,y)\sqrt{\det h(t,y)}.
\end{equation}
In the next subsection, we will analyze cosmological bounces.
\subsection{Bouncing cosmology in flat FRW}\label{sec:bounce-frw}
We consider a spatially flat FRW metric,
\begin{equation}
\mathrm{d}s^{2}
=
-\mathrm{d}t^{2}
+a^{2}(t)\,\delta_{ij}\,\mathrm{d}x^{i}\mathrm{d}x^{j},
\qquad i,j=1,2,3,
\end{equation}
with scale factor $a(t)$. A convenient nonsingular bouncing ansatz, with the bounce at $t=0$, is\footnote{
The averaged Einstein-frame framework discussed in this section can also be applied to other nonsingular bouncing scale factors, including matter-bounce and radiation-bounce profiles~\cite{Brandenberger:1993ef,Novello:2008ra,Ijjas:2018qbo}. Thus, the conclusions do not depend on the particular bouncing ansatz chosen in Eq.~\eqref{eq:bounce-scale-factor}.
} \begin{equation}\label{eq:bounce-scale-factor}
a(t)=a_{\min}\sqrt{1+\left(\frac{t}{t_{0}}\right)^{2}},
\end{equation}
where $a_{\min}\equiv a(0)$ and $t_{0}$ sets the bounce timescale. 
The averaged Einstein-frame analysis extends to any non-singular bounce, since the bounce-point contribution to the four-dimensional NEC violation depends only on the timescale $t_0$ set by $\ddot{a}(0)/a(0)$ and not on the global profile of $a(t)$. 

where $a_{\min}\equiv a(0)$ and $t_{0}$ sets the bounce timescale. This ansatz satisfies,
\begin{equation}
a(0)=a_{\min},\qquad \dot a(0)=0,\qquad \ddot a(0)=\frac{a_{\min}}{t_{0}^{2}}>0.
\end{equation}
Assuming standard four-dimensional Einstein equations with a perfect fluid,
$T^{\mu}{}_{\nu}=\mathrm{diag}(-\rho,p,p,p)$, the Friedmann equations give
\begin{align}
\rho(t) &= \frac{3}{8\pi G_{4}}\,H^{2},\\
p(t) &= -\frac{1}{8\pi G_{4}}\left(2\frac{\ddot a}{a}+H^{2}\right),
\qquad H\equiv \frac{\dot a}{a}.
\end{align}
Hence,
\begin{equation}
\rho+p
=
-\frac{1}{4\pi G_{4}} \dot{H}.
\end{equation}
Evaluated at the bounce,
\begin{equation}
\left.(\rho+p)\right|_{t=0}
=
-\frac{1}{4\pi G_{4}}\frac{\ddot a(0)}{a(0)}
=
-\frac{1}{4\pi G_{4}\,t_{0}^{2}}
<0,
\end{equation}
so the null energy condition is violated at $t=0$.

Equivalently, for the flat FRW background $\bar g_{\mu\nu}$ one finds, for any null vector
$\ell^{\mu}$,
\begin{equation}\label{eq:Ricci-null-null-FRW}
R_{\mu\nu}(\bar g)\,\ell^{\mu}\ell^{\nu}
=
2\left(H^{2}-\frac{\ddot a}{a}\right).
\end{equation}
For the bounce scale factor \eqref{eq:bounce-scale-factor},
\begin{equation}\label{eq:Ricci-null-null-bounce}
R_{\mu\nu}(\bar g)\,\ell^{\mu}\ell^{\nu}
=
\frac{2t^{2}-2t_{0}^{2}}{(t_{0}^{2}+t^{2})^{2}}.
\end{equation}
Therefore the four-dimensional NEC is violated in the interval
\begin{equation}\label{eq:NEC-interval}
-t_{0}<t<t_{0}.
\end{equation}
Combining \eqref{eq:REw-final-factorized-Tn} and \eqref{eq:Ricci-null-null-bounce},
we find that, for $n=6$,
\begin{equation} \label{kushafgan}
\boxed{
\langle R_E\rangle_w(t)
=
\frac{2t^{2}-2t_{0}^{2}}{(t_{0}^{2}+t^{2})^{2}}
+
\frac{1}{2L^{2}(5p^{2}+2)}
\left[
246p^{2}
+
5\left(1-\sqrt{1-4p^{2}}\right)
\right],
\qquad (|p|<\tfrac12).
}
\end{equation}
At the bounce $t=0$, the first term is most negative, therefore we will focus on that limit. We need to check if the second term can be made sufficiently large to make the overall RHS of \eqref{kushafgan} can be made positive. For $p=0$, i.e. in the limit where internal directions are time independent. One can see in such a scenario, it is not possible to make the LHS of \eqref{kushafgan} positive, as required by worldsheet symmetry. Therefore bouncing open or flat FLRW cosmologies are inconsistent with worldsheet string theory\cite{Parikh:2015bja}. For $p\neq0$, i.e. for time-dependent scenarios, it is not the case.
We find that the RHS of \eqref{kushafgan} can be made positive if, \begin{eqnarray}
    S(p)>r^2
\end{eqnarray}
where,
\begin{eqnarray}
    S(p)\equiv\frac{246p^2+5(1-\sqrt{1-4p^2})}{4(5p^2+2)},\qquad r\equiv\frac{L}{t_0}.
\end{eqnarray}

For $L>>t_0$, we see that the LHS
of \eqref{kushafgan} is never positive at the bounce. This indicates the averaged NEC is never satisfied in higher dimensions for any allowed values of $p$.  On the other hand, for $L<<t_0$, $r\rightarrow0$, therefore the RHS of \eqref{kushafgan} is always positive for any $|p|<\frac{1}{2}$, except $p=0$. Interestingly,
the latter limit is the scale separated limit, i.e. a configuration where the size of the external spacetime is much larger than the compact internal space. Therefore, we can see that for any time- dependent compactification within the allowed range $|p|<\frac{1}{2}$, (except $p=0$)
allows a bounce while keeping the LHS positive as required by the worldsheet symmetry eq. \eqref{abdul}. We note that the external four-dimensional Ricci scalar for the bouncing FRW background  evaluates to $R_4(t) = 6/(t_0^2 + t^2)$, which is finite throughout the evolution and reaches its maximum value $R_4(0) = 6/t_0^2$ at the bounce. The curvature scale therefore stays bounded, and our scale separation discussion holds throughout the bounce.
In the case when $r\sim O(1)$, there are regions of
 values for $|p|<\frac{1}{2}$, which can not make the LHS positive, unlike scale
separated limit. \\\\ For any $r$, with $p=0$, i.e.\ for time-independent internal directions, it is not possible to make the LHS of \eqref{kushafgan} positive at the bounce, as required by worldsheet symmetry\cite{Parikh:2015bja}. As we can see, we can clearly circumvent the issue in time-dependent compactification using the averaged Einstein frame condition. Specifically for scale separated solutions, the averaged NEC is always satisfied in higher dimensions for any $|p|<\frac{1}{2}$ (except $p=0$), but allows bouncing in external directions. 
\section{Summary and Discussion}
In this article, we first focused on how worldsheet symmetry leads to the NEC. We considered bosonic string theory, where the string couples to the dilaton, metric, and antisymmetric $B$ field, thereby extending the analysis of \cite{Parikh:2014mja}. Although the presence of the $B$ field makes the computation slightly more involved, it remains straightforward, as shown in Sections~\ref{v} and~\ref{2.1}. The two-dimensional symmetry of the string worldsheet places nontrivial restrictions on admissible spacetime solutions via the NEC, illustrating a deep connection between worldsheet equations and spacetime dynamics. Accordingly, in Sections~\ref{section:unaveraged} and~\ref{bq} we impose the higher-dimensional NEC in the subsequent discussion. We have not included higher-derivative or quantum corrections, which are known to source violations of energy conditions\cite{Bernardo:2021zxo,Faruk:2024usu}. It is well known that objects with negative tension, such as orientifold planes, can violate energy conditions locally\cite{Coudarchet:2023mfs}. However, tadpole cancellation conditions may require that, overall, the energy conditions remain satisfied\cite{Bernardo:2022ony,Kamal:2025qia}. In this work, we instead presented an alternative route to NEC violation in external directions, namely via an averaged condition in a time-dependent setting; see Eq.~\eqref{eq-intXmu}.

In Section~\ref{section:unaveraged}, we studied compactification on a time-dependent compact manifold with a time-independent warp factor (see Eq.~\eqref{eq-metric}). We identified the condition ensuring that the NEC holds in the external directions. A point to note is that this choice keeps Newton's constant time independent, as required. Finally, in Section~\ref{bq}, we explained how Newton's constant can be kept time independent by imposing the averaged condition in Eq.~\eqref{eq-intXmu}. In that section, both the warp factor and the internal coordinates depend on time in periodic fashion. In this way, we obtained a comprehensive understanding of how the NEC in the external spacetime directions emerges under general time-dependent string compactifications. By contrast, if one implements the unaveraged condition and imposes the higher-dimensional NEC pointwise, then the lower-dimensional NEC follows trivially upon compactification. In the usual time-independent compactifications \cite{Parikh:2014mja,Parikh:2015bja}, this unaveraged condition is typically imposed, which guarantees the lower-dimensional NEC from the higher-dimensional one; this observation is often used to argue against bouncing cosmologies.

Bouncing cosmologies are often considered as possible alternatives to the standard Big Bang singularity. To realize a bounce, the universe must transition from contraction to expansion: during contraction the Hubble rate $H$ is negative, while during expansion it is positive. Thus, at the bounce $H$ must increase, passing through $H=0$ and changing sign from negative to positive (see, for example, Eq.~\eqref{eq:bounce-scale-factor}). However, the NEC-violating aspect of such scenarios is often viewed as problematic from the perspective of a string-theory embedding, since worldsheet symmetry enforces the Ricci convergence condition, Eq.~\eqref{abdul}\cite{Parikh:2015bja}. 
In this article we show using \eqref{}
Here   the  NEC violation sourced entirely through the time-dependent internal metric and warp factor contributions. Higher-order corrections do not enter our construction as a source of NEC violation.
We emphasize that this conclusion is not tied to the specific bouncing ansatz used in the explicit computation; the same averaged criterion applies to the entire class of non-singular bouncing FRW geometries, as noted in the footnote on page~12. Most interestingly, the averaged Einstein-frame condition also allows scale-separated solutions.

Scale separation is a key phenomenological requirement: the compact internal space must be much smaller than the non-compact spacetime directions, so that the physics can be reliably captured by a lower-dimensional effective description. 
Such a hierarchy is notoriously difficult to realize in known string-theory vacua\cite{Gautason:2015tig}. But, time-dependent compactification through the averaged Einstein-frame condition provides interesting new avenues to achieve scale separation. We plan to revisit the no-go theorems related to scale separation in future work
in light of time-dependent compactification\cite{Russo:2019fnk,Gautason:2015tig,Andriot:2025cyi,Cribiori:2021djm,Andriot:2023wvg,Rudelius:2022gbz}. More precisely, we would like to test the bounds derived from the top-down compactification scheme discussed in \cite{Gautason:2015tig,Andriot:2025cyi}, under the averaged Einstein-frame condition.

We emphasize that, from the viewpoint of energy conditions, our analysis shows that bouncing cosmologies can be compatible with string theory via time-dependent compactification. Our work is therefore an important first step towards constructing a bouncing cosmology in string theory\cite{Dasgupta:2019gcd,Marconnet:2022fmx}. An important open question is whether, for a given bouncing ansatz, one can solve the full higher-dimensional string equations of motion and determine what fluxes (and other ingredients) are required to support the solution\cite{Bernardo:2021rul,Chatterjee:2015uya}. The string-theoretic techniques developed in \cite{Bernardo:2021zxo,Dasgupta:2019gcd,Bernardo:2021rul,Dasgupta:2018rtp} for solving the equations of motion in the presence of time-dependent fluxes, quantum corrections, and time-dependent internal geometries provide a natural setting in which such a UV completion can be carried out. These works typically employ the unaveraged condition, in which the internal volume is kept strictly constant; extending them to the periodically time-dependent case required by the averaged Einstein-frame condition would furnish an explicit UV realization of the construction presented here, and is the subject of ongoing work. It also remains to clarify the regime of validity, lifetime, and status of any lower-dimensional effective description of these backgrounds\cite{Bernardo:2021zxo,Marconnet:2022fmx}. Finally, it would be particularly interesting to understand how the averaged Einstein-frame constraints relate to swampland conjectures, moduli stabilization, the Trans-Planckian Censorship Conjecture (TCC), and the notion of quantum breaking time\cite{Blumenhagen:2020doa,Russo:2019fnk,Brandenberger:2021pzy,Chatterjee:2015uya,Dvali:2018jhn,Becker:2024ijy,Ahmed:2023uem,Hassfeld:2023kpu,Moghtaderi:2025cns}.

\appendix
\section{Appendix A: Non-negativity of \texorpdfstring{$\text{tr}\left[(h^{-1}(\bar{\nabla}_lh))^2 \right]$}{1}}\label{app-proof-trace}In this section, we prove that
\begin{equation}\label{eq-trace}
\text{tr}\left[(h^{-1}(\bar{\nabla}_l h))^2 \right] \geq 0,
\end{equation}
is non-negative, where we denote $\bar{\nabla}_l\equiv l\cdot \bar{\nabla}$. First, since $h_{pq}$ is a symmetric matrix, we can diagonalize it using matrix notation as:
\beq 
h=R D R^{-1}\,,
\eeq 
where $D$ is a diagonal matrix with non-negative elements (as $h_{pq}$ is a positive semi-definite matrix), and $R$ is an orthogonal matrix (i.e., $R^{-1}=R^t$). Notice that the covariant derivative $\bar{\nabla}_l$ of $h$ can affect both $D$ and $R$, as they may depend on the external coordinates $x^\mu$. Taking this into account, we can write: 
\begin{align}
\bar{\nabla}_l h=&R(\bar{\nabla}_lD+R^{-1}(\bar{\nabla}_l R)D+D(\bar{\nabla}_lR^{-1})R)R^{-1}=R(\bar{\nabla}_lD+[M,D])R^{-1}\,,    
\end{align} 
where we defined $M\equiv R^{-1}\bar{\nabla}_lR=-(\bar{\nabla}_l R^{-1})R$, and the last equality follows from $\bar{\nabla}_l(R^{-1}R)=0$. Note in particular that the orthogonality of $R$ implies that $M$ is antisymmetric (i.e. $M^t=-M$), and this in turn implies that $[M,D]$ is symmetric. 
\\\\This allows us to write \eqref{eq-trace} as
\begin{align}\label{eq-trace-terms}
\text{tr}\left[(h^{-1}(\bar{\nabla}_lh))^2 \right]=&\text{tr}\left[D^{-1}\bar{\nabla}_lDD^{-1}\bar{\nabla}_lD \right]+2~ \text{tr}\left[D^{-1}\bar{\nabla}_lDD^{-1}[M,D]\right]\nonumber\\
&+\text{tr}\left[D^{-1}[M,D]D^{-1}[M,D]\right]\,.
\end{align}
In order to prove that this is non-negative, let us study each term of the RHS in detail: The first term $\text{tr}\left[D^{-1}\bar{\nabla}_lDD^{-1}\bar{\nabla}_lD \right]$ is trivially non-negative as $D$, $D^{-1}$ and $\bar{\nabla}_lD$ are diagonal matrices. The second term vanishes as
\beq 
\text{tr}\left[D^{-1}\bar{\nabla}_lDD^{-1}[M,D]\right]=\text{tr}\left[\bar{\nabla}_lD D^{-1}M\right]-\text{tr}\left[D^{-1}\bar{\nabla}_lDM\right]=\text{tr}\left[[\bar{\nabla}_lD, D^{-1}]M\right]=0\,,
\eeq 
where we used that $\bar{\nabla}_lD$ and $D^{-1}$ are diagonal matrices and thus they commute. Finally, the last term of the RHS of \eqref{eq-trace-terms} can be written as
\begin{align}
\text{tr}\left[D^{-1}[M,D]D^{-1}[M,D]\right]=&\sum_{p,q,r,s} (D^{-1})_{pq}[M,D]_{qr}(D^{-1})_{rs}[M,D]_{sp}\nonumber\\
=&\sum_{p,r}\underbrace{D_{pp}^{-1}D_{rr}^{-1}[M,D]_{pr}[M,D]_{rp}}_{\displaystyle\geq0}\geq0\,,
\end{align}
which is a sum of non-negative terms, as each element $D_{pp}^{-1}$ of the diagonal matrix $D^{-1}$ is non-negative, and $[M,D]_{pr}[M,D]_{rp}=([M,D]_{pr})^2\geq 0$ because $[M,D]$ is a symmetric matrix. This completes the proof of the non-negativity of \eqref{eq-trace-terms}. 
\section{Appendix B: Computation of Ricci Tensor}
We will now express the covariant derivatives $\nabla_\mu h_{pq}$ and $\nabla_\mu\nabla_\nu h_{pq}$ in terms of the covariant derivatives $\bar{\nabla}_\mu$ with respect to the metric $\bar{g}$ without the warp factor using the following relations:
\begin{eqnarray}
    &&\nabla_\alpha h_{pq}=\bar{\nabla}_\alpha h_{pq}=\partial_\alpha h_{pq}\,,\\
&&\nabla_\alpha\nabla_\beta h_{pq}=\bar{\nabla}_\alpha\bar{\nabla}_\beta h_{pq}-\frac{\bar{\nabla}_\alpha\Omega}{\Omega}\bar{\nabla}_\beta h_{pq}-\frac{\bar{\nabla}_\beta\Omega}{\Omega}\bar{\nabla}_\alpha h_{pq}+\bar{g}_{\alpha\beta}\frac{\bar{\nabla}^\sigma\Omega}{\Omega}\bar{\nabla}_\sigma h_{pq}\,,
\end{eqnarray}
First, we can plug in $g_{\mu\nu}(x,y)\equiv\Omega^2(x,y)\bar{g}_{\mu\nu}(x)$ to simplify the last line of eq. \eqref{eq-Ricci-Omegagbar} as
\beq 
\frac{1}{2}g^{\rho\sigma}\nabla_pg_{\rho\alpha}\nabla^pg_{\sigma\beta}-\frac{1}{4}g^{\rho\sigma}\nabla_pg_{\alpha\beta}\nabla^pg_{\rho\sigma}-\frac{1}{2}h^{pq}\nabla_p\nabla_qg_{\alpha\beta}=-\frac{1}{d}\bar{g}_{\alpha\beta}\Omega^{2-d}\nabla^p\nabla_p\Omega^d\,.%-\bar{g}_{\mu\nu}((d-1)\nabla^p \Omega \nabla_p \Omega+\Omega \nabla^p\nabla_p \Omega)\,,
\eeq
Finally, combining all the results, we
 rewrite eq. \eqref{eq-Ricci-Omegagbar} in terms of the metric $\bar{g}$ without the warp factor as
\begin{align}\label{eq-Ricci-gbar12}
{R}_{\alpha\beta}(g^E) &= R_{\alpha\beta}(\bar{g})-(d-2)\frac{\bar{\nabla}_\alpha \bar{\nabla}_\beta \Omega}{\Omega}+2(d-2)\frac{\bar{\nabla}_\alpha \Omega}{\Omega}\frac{\bar{\nabla}_\beta \Omega}{\Omega}\nonumber\\
&\quad \quad-\bar{g}_{\alpha\beta} \left(\frac{\bar{\nabla}^\sigma\bar{\nabla}_\sigma\Omega}{\Omega}+(d-3) \frac{\bar{\nabla}^\sigma \Omega}{\Omega}\frac{\bar{\nabla}_\sigma \Omega}{\Omega}\right)\nonumber\\
&\quad\quad-\frac{1}{4}\bar{\nabla}_{\alpha}h^{pq}\bar{\nabla}_\beta h_{pq} -\frac{1}{2}h^{pq}\bar{\nabla}_\alpha\bar{\nabla}_\beta h_{pq}+\frac{1}{2}\frac{\bar{\nabla}_\alpha\Omega}{\Omega}h^{pq}\bar{\nabla}_\beta h_{pq}+\frac{1}{2}\frac{\bar{\nabla}_\beta\Omega}{\Omega}h^{pq}\bar{\nabla}_\alpha h_{pq}\nonumber\\
&\quad\quad-\frac{1}{2}\bar{g}_{\alpha\beta}\frac{\bar{\nabla}^\sigma\Omega}{\Omega}h^{pq}\bar{\nabla}_\sigma h_{pq}
%\nonumber\\
%&\quad\quad 
-\frac{1}{d}\bar{g}_{\alpha\beta}\Omega^{2-d}\nabla^p\nabla_p\Omega^d\,.
\end{align}\\
Noting that 
\beq 
\bar{\nabla}_\nu h^{pq} \bar{\nabla}_\mu h_{pq}=- h^{pm}\bar{\nabla}_\nu h_{mn} h^{nq}\bar{\nabla}_\mu h_{pq}=-\text{tr}\left(h^{-1}(\bar{\nabla}_\nu h )h^{-1}\bar{\nabla}_\mu h\right)\,,
\eeq 
we can write two of the terms in \eqref{eq-Ricci-gbar12} as 
\beq 
-\frac{1}{2}\text{tr}(h^{-1}\bar{\nabla}_\nu\bar{\nabla}_\mu h)-(d-2)\frac{\bar{\nabla}_\nu\bar{\nabla}_\mu\Omega }{\Omega}=\frac{1}{2}\bar{\nabla}_\nu h^{pq} \bar{\nabla}_\mu h_{pq}-(d-2)\frac{\bar{\nabla}_\nu\Omega }{\Omega}\frac{\bar{\nabla}_\mu\Omega }{\Omega}-\bar{\nabla}_\nu X_\mu\,,
\eeq 
and we can plug this in \eqref{eq-Ricci-gbar12}, yielding
\begin{align}
{R}_{\alpha\beta}(g^E) &= R_{\alpha\beta}(\bar{g})+(d-2)\frac{\bar{\nabla}_\alpha \Omega}{\Omega}\frac{\bar{\nabla}_\beta \Omega}{\Omega}-\frac{1}{4}\text{tr}\left(h^{-1}(\bar{\nabla}_\alpha h )h^{-1}\bar{\nabla}_\beta h\right)\nonumber\\
&+\frac{1}{2}\frac{\bar{\nabla}_\alpha\Omega}{\Omega}\text{tr}(h^{-1}\bar{\nabla}_\beta h)+\frac{1}{2}\frac{\bar{\nabla}_\beta\Omega}{\Omega}\text{tr}(h^{-1}\bar{\nabla}_\alpha h)-\bar{\nabla}_\beta X_\alpha\nonumber\\
&-\bar{g}_{\alpha\beta} \bigg(\frac{\bar{\nabla}^\sigma\bar{\nabla}_\sigma\Omega}{\Omega}+(d-3) \frac{\bar{\nabla}^\sigma \Omega}{\Omega}\frac{\bar{\nabla}_\sigma \Omega}{\Omega}+\frac{1}{2}\frac{\bar{\nabla}^\sigma\Omega}{\Omega}\text{tr}(h^{-1}\bar{\nabla}_\sigma h)
+\frac{1}{d}\Omega^{2-d}\nabla^p\nabla_p\Omega^d\bigg)\,.
\end{align}
We can further use \eqref{eq-def-Xmu} to rewrite the second line of the previous equation as,
\begin{align}\label{eq-Ricci-gbar-withX}
{R}_{\alpha\beta}(g^E) &= R_{\alpha\beta}(\bar{g})-(d-2)\frac{\bar{\nabla}_\alpha \Omega}{\Omega}\frac{\bar{\nabla}_\beta \Omega}{\Omega}-\frac{1}{4}\text{tr}\left(h^{-1}(\bar{\nabla}_\alpha h )h^{-1}\bar{\nabla}_\beta h\right)
+\frac{\bar{\nabla}_\alpha\Omega}{\Omega}X_\beta+\frac{\bar{\nabla}_\beta\Omega}{\Omega}X_\alpha
\nonumber\\
&-\bar{\nabla}_\beta X_\alpha-\bar{g}_{\alpha\beta} \bigg(\frac{\bar{\nabla}^\sigma\bar{\nabla}_\sigma\Omega}{\Omega}+(d-3) \frac{\bar{\nabla}^\sigma \Omega}{\Omega}\frac{\bar{\nabla}_\sigma \Omega}{\Omega}+\frac{1}{2}\frac{\bar{\nabla}^\sigma\Omega}{\Omega}\text{tr}(h^{-1}\bar{\nabla}_\sigma h)
+\frac{1}{d}\Omega^{2-d}\nabla^p\nabla_p\Omega^d\bigg)\,.
\end{align}
\label{appendix B}
\section{Appendix C: Deriving eq. (\ref{eq:general-null-nullX-otp}) from eq. (\ref{eq-Ricci-ll-withX})}
Since $\Omega$ depends only on $t$ (and $y$),
\begin{equation}
\bar\nabla_l\Omega=l^\mu\partial_\mu\Omega=l^t\dot\Omega,
\qquad\Rightarrow\qquad
\frac{\bar\nabla_l\Omega}{\Omega}=l^t\frac{\dot\Omega}{\Omega}=l^t\dot f.
\end{equation}
Hence
\begin{equation}\label{eq:term1}
-(d-2)\left(\frac{\bar\nabla_l\Omega}{\Omega}\right)^2
=-(l^t)^2(d-2)\dot f^{\,2}.
\end{equation}

{Next we evaluate the  trace term.}
Because $h_{mn}=h_{mn}(t,y)$,
\begin{equation}
(\bar\nabla_l h)_{mn}=l^\mu\partial_\mu h_{mn}=l^t\dot h_{mn},
\end{equation}
so
\begin{equation}
h^{-1}(\bar\nabla_l h)=l^t(h^{-1}\dot h)=l^t\mathcal{B}.
\end{equation}
Therefore
\begin{equation}\label{eq:term2}
-\frac14\,\tr\!\left(h^{-1}(\bar\nabla_l h)\,h^{-1}(\bar\nabla_l h)\right)
=-(l^t)^2\frac14\,\tr(\mathcal{B}^2).
\end{equation}
Furthermore,
Using $X_\mu=\bar\nabla_\mu\ln(\Omega^{d-2}\sqrt{\det h})$ and \eqref{eq:def-fk},
\begin{equation}
X_\mu=\bar\nabla_\mu\big((d-2)f+k\big).
\end{equation}
Since $f$ and $k$ depend only on $t$ (and $y$), we have
\begin{equation}
X_t=(d-2)\dot f+\dot k,
\qquad
X_i=0,
\end{equation}
and thus
\begin{equation}\label{eq:Xl}
X_l=l^\mu X_\mu=l^t\big((d-2)\dot f+\dot k\big).
\end{equation}
Combining this with $\bar\nabla_l\Omega/\Omega=l^t\dot f$, we obtain
\begin{equation}\label{eq:term3}
2\,\frac{\bar\nabla_l\Omega}{\Omega}\,X_l
=2(l^t\dot f)\Big(l^t\big((d-2)\dot f+\dot k\big)\Big)
=(l^t)^2\Big(2(d-2)\dot f^{\,2}+2\dot f\,\dot k\Big).
\end{equation}
We compute
\begin{equation}
\bar\nabla_\beta X_\alpha=\partial_\beta X_\alpha-\bar\Gamma^\sigma_{\beta\alpha}X_\sigma.
\end{equation}
Since $X_i=0$, we split
\begin{equation}
l^\alpha l^\beta\bar\nabla_\beta X_\alpha
=
l^t l^\beta\bar\nabla_\beta X_t
+
l^i l^\beta\bar\nabla_\beta X_i.
\end{equation}
For the first piece, $\partial_i X_t=0$ and $\bar\Gamma^\sigma_{\beta t}X_\sigma$ does not contribute, hence
\begin{equation}
l^t l^\beta\bar\nabla_\beta X_t
=
(l^t)^2\partial_t X_t
=
(l^t)^2\big((d-2)\ddot f+\ddot k\big).
\end{equation}
For the second piece, $X_i=0$ implies
\begin{equation}
\bar\nabla_\beta X_i=-\bar\Gamma^t_{\beta i}\,X_t.
\end{equation}
In FRW, the only nonzero Christoffel needed is
\begin{equation}
\bar\Gamma^t_{ji}=a\dot a\,\delta_{ji},
\qquad
\bar\Gamma^t_{ti}=0,
\end{equation}
so
\begin{equation}
l^i l^\beta\bar\nabla_\beta X_i
=
-l^i l^j\,\bar\Gamma^t_{ji}\,X_t
=
-(a\dot a)\,\delta_{ij}l^i l^j\,X_t.
\end{equation}
Using $X_t=(d-2)\dot f+\dot k$, we conclude
\begin{equation}\label{eq:term4}
\bar\nabla_l X_l
=
(l^t)^2\big((d-2)\ddot f+\ddot k\big)
-(a\dot a)\,\delta_{ij}l^i l^j\,\big((d-2)\dot f+\dot k\big),
\end{equation}
and therefore
\begin{equation}\label{eq:term4minus}
-\bar\nabla_l X_l
=
-(l^t)^2\big((d-2)\ddot f+\ddot k\big)
+(a\dot a)\,\delta_{ij}l^i l^j\,\big((d-2)\dot f+\dot k\big).
\end{equation}

\subsection*{Final combination}
Substituting \eqref{eq:term1}, \eqref{eq:term2}, \eqref{eq:term3}, and eq. \eqref{eq:term4minus} into \eqref{eq-Ricci-ll-withX}, and collecting the $(l^t)^2$ terms, yields
\begin{equation}\label{eq:general-null-null-app}
\begin{aligned}
R_{\alpha\beta}(g_E)\,l^\alpha l^\beta
&= R_{\alpha\beta}(\bar g)\,l^\alpha l^\beta \\
&\quad + (l^{t})^{2}\Big[
-(d-2)\,\ddot f
+ (d-2)\,\dot f^{\,2}
+ 2\,\dot f\,\dot k
- \ddot k
-\frac{1}{4}\,\tr(\mathcal{B}^{2})
\Big] \\
&\quad + (a\dot a)\,\delta_{ij}l^{i}l^{j}\,
\Big[(d-2)\,\dot f+\dot k\Big],
\end{aligned}
\end{equation}
which is precisely eq.~\eqref{eq:general-null-nullX-otp}.
%============================================================
\appendix

\acknowledgments

The author thanks Facundo Rost, Alexey Koshelev, Jan Pieter van der Schaar, and Maulik Parikh for useful discussions. MMF acknowledges funding support from the Double First Class Postdoctoral Fellowship and Prof.\ Alexey Koshelev’s start-up fund (2023F0201-000-01).

%% [A] Recommended: using JHEP.bst file
\bibliographystyle{JHEP}
 \bibliography{biblio.bib}
\end{document}